\documentclass[main]{aa} 
\usepackage{xspace}
\usepackage{graphicx}
\usepackage{txfonts}
\usepackage{booktabs}
\usepackage{url}
\usepackage{xcolor}
\usepackage[breaklinks, colorlinks, citecolor=blue, linkcolor=blue]{hyperref}
\usepackage{scalerel}
\usepackage{tikz}

\begin{document} 

\newcommand*{\tabhead}[1]{\multicolumn{1}{c}{\bfseries #1}}

\newcommand{\mystar}{{\Large {\fontfamily{lmr}\selectfont$\star$}}}
\newcommand{\pds}{PDS~70}

\newcommand{\jgrp}{JGR-Planets}
\newcommand{\rsos}{Royal Soc. Open Sci.}
\newcommand{\astrolett}{Astron. Lett.}
\newcommand{\psj}{Planet. Sci. J.}
\newcommand{\jatis}{J. Astron. Telesc. Instrum. Syst.}

\newcommand{\mgtwo}{Mg\,II}
\newcommand{\ctwo}{C\,II}
\newcommand{\molh}{H$_2$}
\newcommand{\lyalph}{Ly-$\alpha$}
\newcommand{\fetwo}{Fe\,II}

\newcommand{\lognh}{$\log(N_{\rm H~I})$}
\newcommand{\oi}{[O\,I]}
\newcommand{\suii}{[S\,II]}
\newcommand{\nii}{[N\,II]}
\newcommand{\nai}{Na\,I}
\newcommand{\caii}{Ca\,II}
\newcommand{\mdotacc}{$\dot{M}_{\rm acc}$}
\newcommand{\mdotwind}{$\cdot{M}_{\rm wind}$}
\newcommand{\hei}{He\,I}
\newcommand{\lyalpha}{Ly-$\alpha$}
\newcommand{\halpha}{H$\alpha$}
\newcommand{\mgi}{Mg\,I}
\newcommand{\sii}{Si\,I}
\newcommand{\water}{H$_2$O}
\newcommand{\methane}{CH$_4$}
\newcommand{\cotwo}{CO$_2$}
\newcommand{\htwo}{H$_2$}


\newcommand{\rosat}{\emph{Rosat}}
\newcommand{\galex}{\emph{GALEX}}
\newcommand{\tess}{\emph{TESS}}
\newcommand{\plato}{\emph{PLATO}}
\newcommand{\gaia}{\emph{Gaia}}
\newcommand{\ktwo}{\emph{K2}}
\newcommand{\jwst}{\emph{JWST}}
\newcommand{\kepler}{\emph{Kepler}}
\newcommand{\corot}{\emph{CoRoT}}
\newcommand{\hipp}{\emph{Hipparcos}}
\newcommand{\spitzer}{\emph{Spizter}}
\newcommand{\herschel}{\emph{Herschel}}
\newcommand{\hst}{\emph{HST}}
\newcommand{\wise}{\emph{WISE}}
\newcommand{\swift}{\emph{Swift}}
\newcommand{\chandra}{\emph{Chandra}}
\newcommand{\xmm}{\emph{XMM-Newton}}

\newcommand{\twa}{TW Hydra}
\newcommand{\bpic}{$\beta$~Pictoris}
\newcommand{\abdor}{AB~Doradus}
\newcommand{\rup}{Ruprecht~147}
\newcommand{\etacha}{$\eta$\,Chamaeleontis}
\newcommand{\usco}{Upper~Sco}
\newcommand{\rhooph}{$\rho$~Oph}

\newcommand{\doar}{DoAr\,25}
\newcommand{\epcha}{EP\,Cha}
\newcommand{\rylup}{RY\,Lup}
\newcommand{\hdtwofour}{HD\,240779}

\newcommand{\sigrv}{$\sigma_{\rm RV}$}

\newcommand{\msunyr}{$\rm{M_{\sun} \, yr^{-1}}$}
\newcommand{\etal}{\mbox{\rm et al.~}}
\newcommand{\ms}{\mbox{m\,s$^{-1}~$}}
\newcommand{\kms}{\mbox{km\,s$^{-1}~$}}
\newcommand{\ks}{\mbox{km\,s$^{-1}~$}}
\newcommand{\kse}{\mbox{km\,s$^{-1}$}}
\newcommand{\mse}{\mbox{m\,s$^{-1}$}}
\newcommand{\msy}{\mbox{m\,s$^{-1}$\,yr$^{-1}~$}}
\newcommand{\msye}{\mbox{m\,s$^{-1}$\,yr$^{-1}$}}
\newcommand{\msun}{M$_{\odot}$}
\newcommand{\msune}{M$_{\odot}$}
\newcommand{\rsun}{R$_{\odot}$}
\newcommand{\lsun}{L$_{\odot}~$}
\newcommand{\rsune}{R$_{\odot}$}
\newcommand{\mjup}{M$_{\rm JUP}~$}
\newcommand{\mjupe}{M$_{\rm JUP}$}
\newcommand{\msat}{M$_{\rm SAT}~$}
\newcommand{\msate}{M$_{\rm SAT}$}
\newcommand{\mnep}{M$_{\rm NEP}~$}
\newcommand{\mnepe}{M$_{\rm NEP}$}
\newcommand{\mearth}{$M_{\oplus}$}
\newcommand{\mearthe}{$M_{\oplus}$}
\newcommand{\rearth}{$R_{\oplus}$}
\newcommand{\rearthe}{$R_{\oplus}$}
\newcommand{\rjup}{R$_{\rm JUP}~$}
\newcommand{\msinie}{$M_{\rm p} \sin i$}
\newcommand{\vsinie}{$V \sin i$}
\newcommand{\mbsini}{$M_b \sin i~$}
\newcommand{\mcsini}{$M_c \sin i~$}
\newcommand{\mdsini}{$M_d \sin i~$}
\newcommand{\chisq}{$\chi_{\nu}^2$}
\newcommand{\chinu}{$\chi_{\nu}$}
\newcommand{\chinusq}{$\chi_{\nu}^2$}
\newcommand{\arel}{$a_{\rm rel}$}
\newcommand{\feh}{\ensuremath{[\mbox{Fe}/\mbox{H}]}}
\newcommand{\rphk}{\ensuremath{R'_{\mbox{\scriptsize HK}}}}
\newcommand{\lrphk}{\ensuremath{\log{\rphk}}}
\newcommand{\cs}{$\sqrt{\chi^2_{\nu}}$}
\newcommand{\etaearth}{$\mathbf \eta_{\oplus} ~$}
\newcommand{\etaearthe}{$\mathbf \eta_{\oplus}$}
\newcommand{\searth}{$S_{\bigoplus}$}
\newcommand{\mdotyr}{$M_{\odot}$~yr$^{-1}$}
\newcommand{\micron}{$\mu$m}

\newcommand{\sini}{\ensuremath{\sin i}}
\newcommand{\msini}{\ensuremath{M_{\rm p} \sin i}}
\newcommand{\mplsini}{\ensuremath{\mpl\sin i}}
\newcommand{\teff}{\ensuremath{T_{\rm eff}}}
\newcommand{\teq}{\ensuremath{T_{\rm eq}}}
\newcommand{\logg}{\ensuremath{\log{g}}}
\newcommand{\vsini}{\ensuremath{v \sin{i}}}
\newcommand{\ebv}{E($B$-$V$)}

\newcommand{\Kepler}{\emph{Kepler}~}
\newcommand{\Keplere}{\emph{Kepler}}
\newcommand{\blender}{{\tt BLENDER}~}

\newcommand{\kp}{\ensuremath{\mathrm{Kp}}}
\newcommand{\rstar}{\ensuremath{R_\star}}
\newcommand{\mstar}{\ensuremath{M_\star}}
\newcommand{\loggstar}{\ensuremath{\logg_\star}}
\newcommand{\lstar}{\ensuremath{L_\star}}
\newcommand{\astar}{\ensuremath{a_\star}}
\newcommand{\loglstar}{\ensuremath{\log{L_\star}}}
\newcommand{\rhostar}{\ensuremath{\rho_\star}}

\newcommand{\rp}{\ensuremath{R_{\rm p}}}
\newcommand{\rpl}{\ensuremath{r_{\rm P}}~}
\newcommand{\rple}{\ensuremath{r_{\rm P}}}
\newcommand{\mpl}{\ensuremath{m_{\rm P}}~}
\newcommand{\lpl}{\ensuremath{L_{\rm P}}~}
\newcommand{\rhopl}{\ensuremath{\rho_{\rm P}}~}
\newcommand{\loggpl}{\ensuremath{\logg_{\rm P}}~}
\newcommand{\logrpl}{\ensuremath{\log r_{\rm P}}~}
\newcommand{\logrple}{\ensuremath{\log r_{\rm P}}}
\newcommand{\loga}{\ensuremath{\log a}~}
\newcommand{\logmpl}{\ensuremath{\log m_{\rm P}}~}

\newcommand{\samunits}{$R_{\odot}^2{\rm day}^{-2}$}
\newcommand{\fludensunits}{ergs s$^{-1}$ cm$^{-2}$ \AA$^{-1}$}

\newcommand{\rr}{$\mathcal{R}$}

\title{The diversification and dissipation of protoplanetary disks}

\author{
Eric~Gaidos\inst{\ref{hawaii},\ref{vienna}\thanks{gaidos@hawaii.edu}} 
\and
Lukas~Gehrig\inst{\ref{vienna}} 
\and
Manuel~G\"{u}del\inst{\ref{vienna}}
}
   
\institute{
Department of Earth Sciences, University of Hawai'i at M\={a}noa, Honolulu, Hawai'i 96822 USA\label{hawaii}
\and
Institute for Astrophysics, University of Vienna, 1180 Vienna, Austria\label{vienna} 
}
\date{Received ; accepted }
\abstract 
{Protoplanetary disk evolution exhibits trends with stellar mass, but also diversity of structure, and lifetime, with implications for planet formation and demographics.  We show how varied outcomes can result from evolving structures in the inner disk that attenuate stellar soft X-rays that otherwise drive photoevaporation in the outer disk.  The magnetic truncation of the disk around a rapidly rotating T Tauri star is initially exterior to the corotation radius and ``propeller" accretion is accompanied by an inner magnetized wind, shielding the disk from X-rays.  Because rotation varies little due to angular momentum exchange with the disk, stellar contraction causes the truncation radius to migrate inside the corotation radius, the inner wind to disappear, and photoevaporation to erode a gap in the disk, accelerating its dissipation.  This X-ray attenuation scenario explains the trend of the longer lifetime, reduced structure, and compact size of disks around lower-mass stars.  It also explains an observed lower bound and scatter in the distribution of disk accretion rates.  Disks that experience early photoevaporation and form gaps can efficiently trap solids at a pressure bump at 1--10 au, triggering giant planet formation, while those with later-forming gaps or indeed no gaps form multiple smaller planets on close-in orbits, a pattern that is consistent with observed exoplanet demographics.}    


   \maketitle
%

\section{Introduction}
\label{sec:intro}

The lifetime and structure of a protoplanetary disk (PPD) set the tempo and conditions for planet formation.  The gas disk drives the starward drift and concentration of solids and dampens planetesimal motion, accelerating planet formation \citep{Birnstiel2024}.  It exerts torques that cause the migration of nascent planets \citep{Nelson2018}.  Radial structures in a disk are important in this context \citep[e.g.,][]{Lau2022}.  Observations of young cluster stars show that disk lifetime decreases with increasing stellar mass \citep{Bayo2012,Ribas2015} but with an order-of-magnitude scatter \citep{Pfalzner2024}.  There are exceptionally long-lived disks ($>$10 Myr) around some very low-mass stars \citep[VLMSs; $\lesssim 0.2$\msun;][]{Murphy2018,Flaherty2019,Silverberg2020,Gaidos2022c}.  The size of dust disks and the occurrence of disk structure (cavities or gaps) positively correlate with stellar mass \citep{vanderMarel2021,Kurtovic2021}, although the latter could be in part a selection effect brought about by the former.

While disks can be truncated and their dissipation hastened by close companion stars \citep{Kraus2012a,Akeson2019,Zurlo2020,Zurlo2021} and far-ultraviolet radiation from neighboring O and B stars \citep{Winter2022}, these factors seem to only be important for a small fraction of stars \citep[e.g,][]{Zurlo2021,MannR2015}.  In the canonical scenario for single and wide binary stars, disk lifetime is thought to be set by a competition between viscous or magnetohydrodynamic (MHD) wind-driven accretion versus ``internal" photoevaporation (PE) driven by high-energy emission from the central star \citep{Armitage2011}, specifically soft X-rays \citep{Ercolano2017,Ercolano2021,Ercolano2023,Lin2024}.  When the accretion rate falls below the local PE rate, a gap appears; this occurs at a fraction of the gravitational radius \citep[a few au,][]{Liffman2003}.  The flow of gas into the inner disk and onto the star then declines or halts \citep{Drake2009} and a cavity develops.  Depletion of dust in the inner disk leads to a concomitant decline in infrared emission that is the signature of ``transition" disks \citep{Ercolano2017}.  

Incident stellar radiation on the PPD, including X-rays, can be attenuated by intervening structures in the inner disk.  Many T Tauri stars (TTSs) exhibit optical variability due to occulting dust in accretion streams \citep{Bouvier2003}, winds \citep{Tambovtseva2008}, or instabilities near the corotation radius \citep{Ansdell2016a,Roggero2021}.  ``See-saw" variability in infrared emission has been explained by shadowing of the disk by interior structures \citep{Muzerolle2009,Espaillat2011,Flaherty2012,Fernandes2018,Gaidos2024a}.  

For an interstellar-like composition \citep[e.g.,][]{Guver2009}, the neutral hydrogen column density of the gas corresponding to the observed extinction would be $10^{21}-10^{22}$\,cm$^{-2}$, but some measurements find much higher values \citep{Guedel2005, Guedel2007c,Robrade2007,Schneider2015b,Guenther2018}.  This intervening gas can absorb X-rays from the star that would otherwise reach the disk.  Accreting TTSs have systematically lower observed X-ray emission than their non-accreting coeval counterparts due to circumstellar gas, an effect that is correlated with the accretion rate \citep{Telleschi2007,Guedel2007b,Bustamante2016}.   Some of the most extreme accretors (e.g., FU Orionis-type objects) host dust-depleted gas that is optically thick to 0.1--1~keV X-rays \citep{Guedel2008, Skinner2010, Liebhart2014}.  Attenuation along shallower lines of sight to the disk surface (elevations of a few degrees) could be even higher.      

\citet{Guedel2010} proposed that X-ray-absorbing gas in the immediate environment of accreting TTSs means that only $\gtrsim$1 keV X-rays, not more readily absorbed extreme-ultraviolet (EUV) radiation, reach the outer disk.  \citet{Pascucci2020} suggest that absorption of soft X-rays by a magnetized inner disk wind (IDW) at $\ll$1 au) could suppress any PE wind at $\gtrsim$ a few astronomical units.  The appearance of an IDW can, in turn, depend sensitively on stellar rotation, the large-scale stellar magnetic field, and inner disk accretion \citep{Romanova2015}. This motivates the question of whether this link between conditions around the central star and PE-triggered disk dissipation could explain the disparate fates of disks and trends of the disk lifetime and structure.  Here, we combine an analytical model of disk evolution with models of pre-main-sequence (pre-MS) stars to investigate this possibility.

\section{Model}
\label{sec:model}

In the inner disk, gas at $>1000$K  is partially ionized and influenced by the large-scale stellar magnetic field.  Interior to the corotation radius 
\begin{equation}
\label{eqn:rcor}
    R_{\rm cor} = \left(\frac{GM_{\star}P_{\star}^2}{4\pi^2}\right)^{1/3},
\end{equation} 
where $G$ is the gravitational constant, $M_{\star}$ is the stellar mass, and $P_{\star}$ is the rotation period, the field exerts a negative torque on any disk gas, driving accretion onto the star.  Beyond the corotation radius, the torque is positive and disk gas can be accelerated and ejected along magnetic field lines.  An ionized disk is expected to be truncated where magnetic pressure exceeds the ram pressure of inflowing gas.  Three-dimensional MHD simulations \citep{Takasao2022,Zhu2024} support a relation between the truncation radius, $R_{\rm mag}$, the strength of the large-scale dipole magnetic field at the stellar surface, $B_{\star}$, and the disk accretion rate, \mdotacc\ , originally derived for spherical flow onto a dipole field \citep{Ghosh1979a}:
\begin{equation}
\label{eqn:rmag}
    R_{\rm mag} = \left(\frac{B_{\star}^4R_{\star}^{12}}{2GM_{\star}\dot{M}_{\rm acc}^2}\right)^{1/7}.
\end{equation}
$R_{\rm mag}$ is also where any magnetic field advected inward by disk accretion will accumulate, increasing magnetic pressure.

The structure of the innermost disk and flow in it depends sensitively on the ratio \rr$ \equiv R_{\rm mag}/R_{\rm cor}$.  If $B_{\star}$ is in kilogauss, $R_{\star}$ and $M_{\star}$ are in solar units, $P_{\star}$ is in days, and \mdotacc\ is in units of $10^{-9}$\,\msun\,yr$^{-1}$, 
\begin{equation}
\label{eqn:ratio}
\mathcal{R} = 1.8 B_{\star}^{4/7} R_{\star}^{12/7} \dot{M}_{-9}^{-2/7} M_{\star}^{-10/21} P_{\star}^{-2/3}
.\end{equation}
If $\mathcal{R} < 0.8$ then magnetospheric accretion occurs in an unsteady flow that remains close to the disk midplane.  At intermediate $\mathcal{R}$ values of 0.8--1, magnetospheric accretion takes the form of a highly non-axisymmetric ``funnel" flow out of the midplane and along field lines onto the star \citep{Bouvier2007b}.   However, if $\mathcal{R} \ge 1$, ``propeller accretion" occurs; some disk gas is accelerated in a magneto-centripetal IDW, and the remainder is accreted onto the star \citep{Romanova2015}, with the ejection/accretion ratio, $f$, depending on loading of disk gas onto stellar field lines and the wind's ratio of Alfven to launch radius.  Evidence for such winds includes highly broadened forbidden lines of \ion{O}{I} \citep{Fang2023b}.   The critical \mdotacc\ at which $R_{\rm mag}$ and $R_{\rm cor}$ intersect ($\mathcal{R} = 1$), in other words the transition between magnetospheric and propeller accretion, is:
\begin{equation}
\label{eqn:mdotcrit}
\dot{M}_{\rm crit} = 8.2 \times 10^{-9} B_{\star}^2 R_{\star}^6 M_{\star}^{-5/3} P_{\star}^{-7/3} M_{\odot} {\rm yr}^{-1}.    
\end{equation}

The X-ray emission from the star is calculated using  the standard broken power law dependent on the Rossby number, $Ro = P_{\star}/\tau_c$ \citep{Wright2011}, where $\tau_c$,  the convective turnover time in the stellar interior, is taken to be $12.6 L_{\star}^{-1/2}$\,days and $L_{\star}$ is the stellar luminosity in solar units \citep{Jeffries2011}.
\begin{equation}
\label{eqn:lx}
L_{\rm X} = \frac{7.4 \times 10^{-4} L_{\star}}{1 + \left(Ro/0.13\right)^{2.18}}.
\end{equation}

A geometrically thin wind\footnote{Winds may actually be more radially extended \citep[e.g.,][]{Pascucci2024} but to first order this does not affect the column density.} launched from both sides of a disk at $R_{\rm mag}$ and an angle, $\theta$, with respect to the disk normal will impose an atomic column density, $N_H$, between the star and disk of 
\begin{equation}
    \label{eqn:coldens1}
    N_{\rm H} \sim \frac{f \dot{M}_{\rm acc}}{4\pi R_{\rm mag} \cos \theta \sqrt{\mu \gamma m_p k_B T}}, 
\end{equation}
where we assume that lines of sight to the disk pass through the transonic flow near the base of the wind, and where $\mu = 2.38$, $\gamma= 1.4$, and $T$ are, respectively, the molecular weight, adiabatic constant, and temperature of the flow.  We take $T$ to be that of a passively heated inner disk wall:  $\approx T_{\star}\sqrt{R_{\star}/R_{\rm trunc}}$ \citep{Dullemond2001}.  The ejection/accretion ratio, $f$, is taken to be 0.2, near typical values inferred by \citet{WatsonD2016} and \citet{Serna2024}.  This yields 
\begin{equation}
\label{eqn:coldens2}
N_H \approx 4 \times 10^{21} {\rm cm}^{-2} \dot{M}_{-9}^{17/14} R_{\star}^{-43/28} M_{\star}^{3/28}B_{\star}^{-3/7} (T_{\star}/4000)^{1/4}.
\end{equation}

We calculated the photoelectric absorption of 0.1--1\,keV X-rays by intervening solar-metallicity gas with $N_H$ using the {\tt XSPEC} package \citep{Arnaud1996}.  We modeled coronal X-ray emission to be a two-temperature (0.5 and 2 keV), equal emission measure plasma like that typically used to describe TTSs \citep[e.g.,][]{Telleschi2007}, and we adopted coronal abundances found in Taurus stars \citep{Guedel2007}.  Unit optical depth is reached at $N_H \approx 10^{21}$~cm$^{-2}$, weakly dependent on the coronal X-ray spectrum (Appendix \ref{sec:sensitivity}).   

We used the relation between PE wind mass loss, $\dot{M}_{\rm PE}$, and (attenuated) stellar 0.1--1 keV  X-ray luminosity, $L_X$ derived from 2D MHD simulations \citep[Eqn. 9 in ][]{Ercolano2021}.  Their relation is nonlinear and convex with $L_X$ due to self-shielding by the PE wind itself \citep[see also][]{Gorti2009}.  At $L_x < 10^{28}$\,ergs~sec$^{-1}$, we used a matched linear relation, $\dot{M}_{\rm PE} = 2.56 \times 10^{-10} (L_X/10^{28})$\,\msun\,yr$^{-1}$.

We adopted Dartmouth stellar evolution models for pre-MS values of $R_{\star}$, $T_{\star}$, and $L_{\star}$ \citep[][see our Appendix \ref{sec:stellar_model} for an investigation into the effects of model choice]{Dotter2008,Feiden2015}.  We used the self-similar solution for time-dependent accretion in a viscous disk with a uniform viscosity \citep{Linden-Bell1974,Hartmann1998}:  
\begin{equation}
\label{eqn:accrete}
    \dot{M}_{\rm acc} = \dot{M}_0 \left(1 + t/t_0\right)^{-5/4}.
\end{equation}
The initial accretion rate was taken to be $\dot{M}_0 = 2.5\times 10^{-7} M_{\star}^{1.6}$ \msun~yr$^{-1}$, based on observations of ``flat-spectrum" young stellar objects considered to be in the Class I--II transition \citep[Eqn. 3 in ][]{Gehrig2023} and a viscous timescale of $t_0 = 1 \times 10^5$ yr.

For given star and prescribed \mdotacc, the remaining parameters are $B_{\star}$ and $P_{\star}$.  We adopted a nominal value for $B_{\star}$ of 0.44\,kG as 22\% of the total field strength \citep{Lavail2019} of 2 kG, typical for magnetically saturated stars \citep{Reiners2022}, but considered a range of 0.1--1\,kG \citep[see Fig. 9 in ][]{Gehrig2022}.  Although protostars should spin up as they contract, observations suggest that the $P_{\star}$ of Class I and II objects remain relatively constant due to torques from the disk \citep[see also the discussion in \citealt{Mueller2024}]{Koenigl1991,Ostriker1995}.   We assume that $P_{\star}$ is fixed to a value between 1 and 10 days while a disk is present \citep{Smith2023,Serna2024}, but consider different scalings of $P_{\star}$ with other parameters in Appendix \ref{sec:sensitivity}.
  
\section{Results}
\label{sec:results}

\subsection{Fiducial and solar-mass scenarios}

At the transition to the Class II phase, low-mass protostars are still inflated but already rapidly rotating \citep[$P_{\star} \sim 1-10$ days,][]{Smith2023}.  As a result, \mdotacc\ is initially below $\dot{M}_{\rm crit}$, $\mathcal{R} > 1$, and for typical $P_{\star}$ and $B_{\star}$ (see below), accretion will be accompanied by an IDW.  Based on Eqn. \ref{eqn:coldens2}, for solar-mass stars the optical depth to 0.1--1 keV X-rays will be $\sim$300 and the disk is completely shielded at the beginning of the Class II phase.\footnote{Infall during the Class I phase will shield the disk from X-rays at earlier times, regardless of the inner disk accretion mode.}  If $P_{\star}$ is nearly constant due to the star's exchange of angular with the disk, then the pre-MS star and its magnetic field will contract with time, $\dot{M}_{\rm crit}$ will decline faster than \mdotacc, and $\mathcal{R}$ will decrease.  Eventually, $\mathcal{R}$ reaches unity, disk accretion enters the magnetospheric regime, the IDW disappears, and stellar X-rays can reach the disk unimpeded.   If $\dot{M}_{\rm PE}$ also exceeds \mdotacc\ then PE opens a gap in the disk (at time $t_{\rm gap}$), causing the disk to more rapidly dissipate.  The dependence of $M_{\rm crit}$ on $B_{\star}$ and inverse dependence on $P_{\star}$ (Eqn. \ref{eqn:mdotcrit}) means that disks around more rapidly rotating stars with stronger fields will tend to be longer-lived.       

\begin{figure}[h!]
    \centering
        \includegraphics[width=0.49\textwidth]{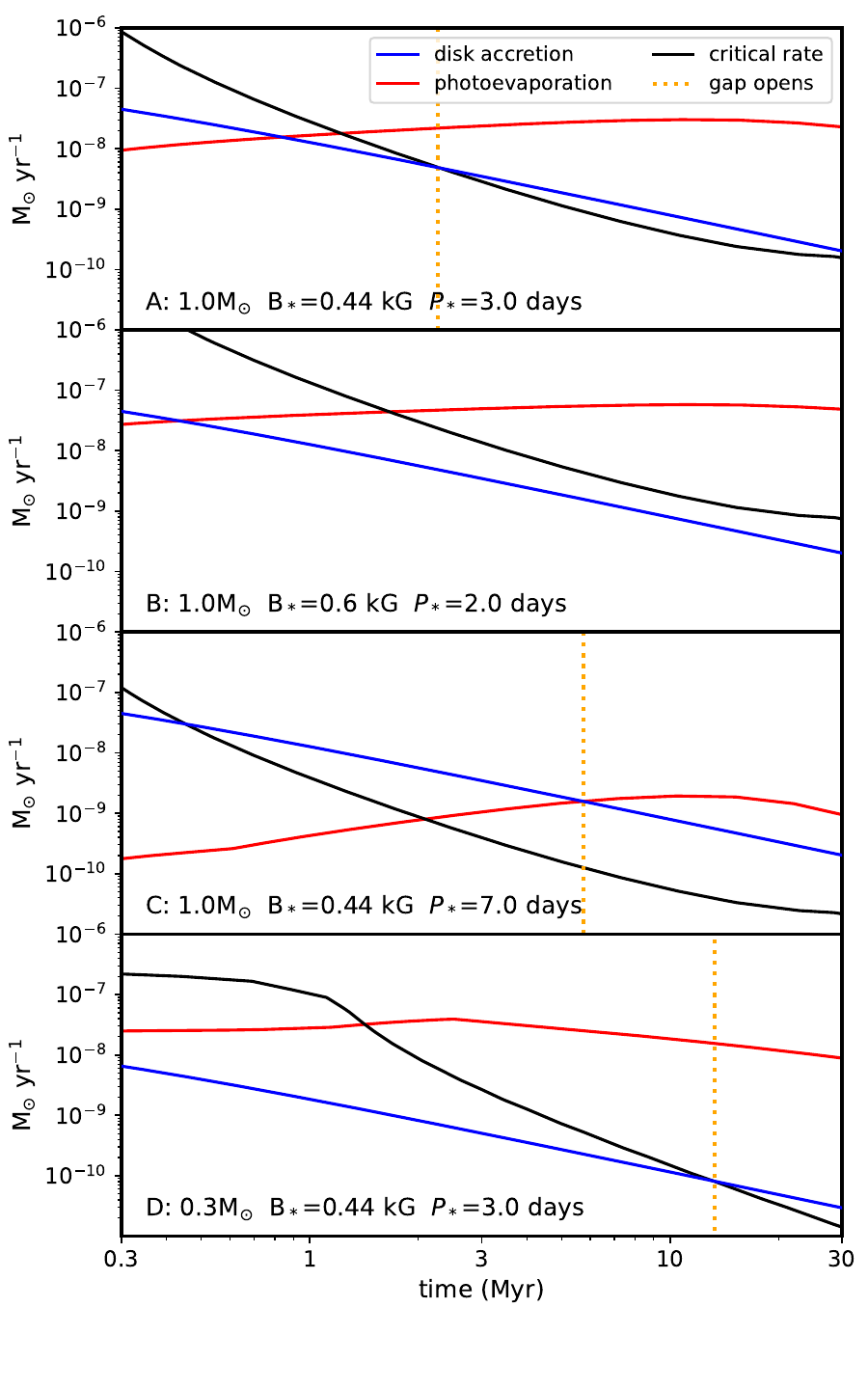}
    \caption{Four scenarios of disk evolution. A: With $M_{\star} = 1$\,\msun, $B_{\star} = 0.44$\,kG, and $P_{\star} = 3$\,days, \mdotacc\ exceeds $\dot{M}_{\rm crit}$ at 2.3\,Myr, $R_{\rm mag}$ retreats inside $R_{\rm cor}$, an IDW ceases, and stellar X-rays open a disk gap.  B: In the case of a more rapidly rotating ($P_{\star}=2$\,days) star with a stronger (0.6\,kG) field,  \mdotacc\ never exceeds $\dot{M}_{\rm crit}$ and no gap opens.  C:  For a more slowly rotating ($P_{\star} = 7$\,day) star, there is no IDW but lower $L_X$ and PE delays $t_{\rm gap}$ to 5.7\,Myr.  D: In the 0.3\msun\ case, lower \mdotacc\ remains below $dot{M}_{\rm crit}$ for longer, delaying $t_{\rm gap}$ to 13.3\,Myr.}
    \label{fig:evolution}
\end{figure}

\begin{figure}[h!]
    \centering
        \includegraphics[width=0.49\textwidth]{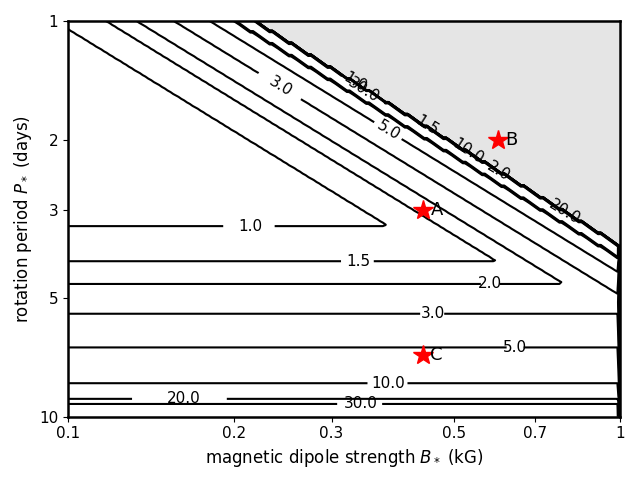}
\caption{$t_{\rm gap}$ (Myr) as a function of $B_{\star}$ and $P_{\star}$ for $M_{\star} = 1$\msun.  Labeled stars correspond to scenarios A, B, and C in Fig. \ref{fig:evolution}  In the shaded region, \mdotacc\ remains below $\dot{M}_{\rm crit}$ and no gap opens for the entire simulation.}
    \label{fig:evolution2}
\end{figure}

Figure \ref{fig:evolution} plots \mdotacc, the $\dot{M}_{\rm PE}$, and the critical accretion rate at which $\mathcal{R} = 1$ for four scenarios.  Time is Dartmouth stellar model time in millions of years.\footnote{For a discussion of the complexities of pre-MS model time for young stellar objects, see Sect. 4 of \citet{Doppmann2005}.} In Scenario A ($M_{\star} = 1$\,\msun, $B_{\star} = 0.44$\,kG and $P_{\star} = 3$\,days), $\dot{M}_{\rm crit}$ falls below \mdotacc, the IDW ceases, and a gap opens at $t_{\rm gap} = 2.3$\,Myr.   In scenario B, with a stronger field ($B_{\star}=0.6$\,kG) and faster rotation ($P_{\star}=2$\,days), \mdotacc\ remains below $\dot{M}_{\rm crit}$ for the entire simulation, and the IDW persists.  A gap could eventually open once \mdotacc\ is insufficient for the wind to absorb most soft X-rays (see Sect. \ref{sec:threshold}).  In Scenario C, the slower rotation ($P_{\star}=7$\,days) causes the IDW to disappear by 0.5\,Myr, but X-ray emission is also lower due to slow rotation and more time elapses (5.7\,Myr) before \mdotacc\ falls below $\dot{M}_{\rm PE}$.  Scenario D is the same as A except for a lower-mass (0.3\,\msun) star; PE is suppressed by an IDW for longer (13.3\,Myr), mostly due to lower \mdotacc.  For a given $M_{\star}$ and \mdotacc\ history, $t_{\rm gap}$ depends only on $B_{\star}$ and $P_{\star}$.  Figure \ref{fig:evolution2} plots this dependence for $M_{\star} = 1$\msun\ over the ranges $B_{\star} \in [0.1,1]$\,kG and $P_{\star} \in [1,10]$\,days, showing the locations of Scenarios A, B, and C in this parameter space.  

\subsection{Disk lifetime and stellar mass}
\label{sec:lifetime}

\begin{figure}[h!]
    \centering
        \includegraphics[width=0.49\textwidth]{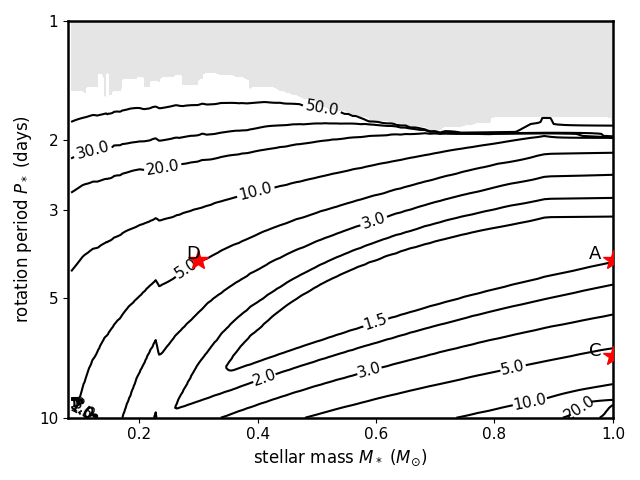}
\caption{Same as Fig. \ref{fig:evolution2}, but with $t_{\rm gap}$ (Myr) as a function of $M_{\star}$ and $P_{\star}$ for $B_{\star} = 0.44$ kG.  Stars correspond to scenarios A, C, and D in Fig. \ref{fig:evolution}.}
    \label{fig:evolution3}
\end{figure}

For fixed $M_{\star} = 1$\,\msun\ but varying $B_{\star}$ and $P_{\star}$, the gap opening time, $t_{\rm gap}$, spans more than an order of magnitude (Fig.\ref{fig:evolution2}), consistent with the variation inferred from disk statistics \citep{Pfalzner2024}.   Superposed on this variation is a trend of longer disk life with lower stellar mass.  This pattern is largely due to the correlation between \mdotacc\ and $M_{\star}$, which in turn is based on observations of flat-spectrum protostars \citep{Gehrig2023} but which is also widely observed among TTSs \citep{Manara2023}.  A longer elapsed time is required for $M_{\rm crit}$ to decline to reach a lower \mdotacc\ and the IDW to disappear (compare scenarios A and D in Fig. \ref{fig:evolution}).  Figure \ref{fig:evolution3} shows that $t_{\rm gap}$ systematically increases with lower $M_{\star}$ for a fixed $B_{\star} = 0.44$kG, except for the slowest rotators.  This is consistent with observations showing longer disk lifetimes around lower-mass stars \citep{Bayo2012,Ribas2015,Silverberg2020}.  $t_{\rm gap}$ increases with decreasing $P_{\star}$: the tendency of VLMS TTSs to be more rapidly rotating \citep[e.g.,][]{Herbst2002,Kounkel2022,Smith2023} means that disks around these objects will be even longer lived.  The strong $R_{\star}$ and weaker $M_{\star}$ dependence of Eqn. \ref{eqn:coldens2} also indicate that, at a given \mdotacc, a more compact IDW around a lower-mass star will more strongly attenuate X-rays.

For a more realistic treatment of the star-disk angular momentum exchange that affects $P_{\star}$, and thus disk dissipation, we performed disk evolution simulations using the self-consistent hydrodynamic model described in \citealt{Gehrig2022,Cecil2024}.  This includes star-disk torques as well as the effects of a PE wind.  The results are described in Appendix \ref{sec:numerical} and support the dependence of $t_{\rm gap}$ on $M_{\star}$, $P_{\star}$, and $B_{\star}$ illustrated by Figs. \ref{fig:evolution2} and \ref{fig:evolution3}.

\section{Discussion}
\label{sec:discussion}

\subsection{Threshold accretion rate}
\label{sec:threshold}
\begin{figure}
    \centering
    \includegraphics[width=0.49\textwidth]{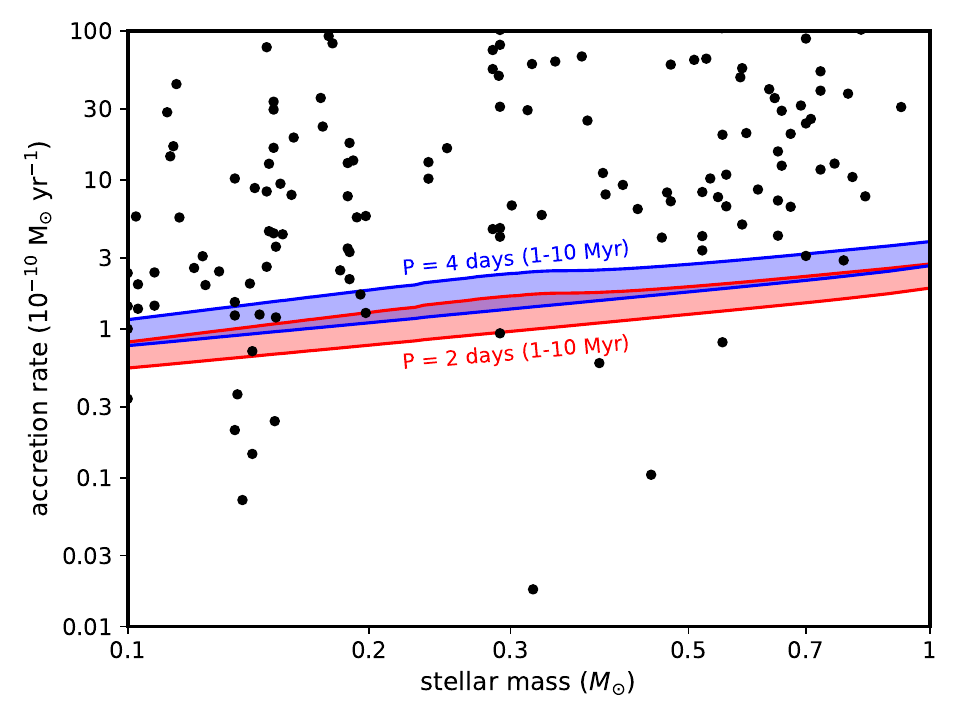}
    \caption{Threshold accretion rate below which a wind cannot effectively shield the disk from PE-driving X-ray radiation, vs. stellar mass.  Curves are calculations for different rotation periods and ages of 1--10 Myr, and points are values of \mdotacc\ from \citet{Manara2023}.}
    \label{fig:threshold}
\end{figure}

Decline in \mdotacc\ with time means that the gas column in any IDW (see Eqn. \ref{eqn:coldens2}) will eventually be insufficient ($N_H \ll 10^{21}$\,cm$^{-2}$) to shield the disk from stellar X-rays.  Thus, PE will inevitably trigger accelerated dissipation of the disk and rapid decline in \mdotacc.  This effect should appear as a baseline \mdotacc\ level, below which few accreting TTSs are found even in sensitive surveys.  \citet{Thanathibodee2023} report that the lowest T Tauri accretors have $\sim 10^{-10}$\,\msun\,yr${_1}$, well above the detection limit of $1-5 \times 10^{-11}$\,\msun\,yr${_1}$.  We calculated the baseline by assuming $R_{\rm mag} = R_{\rm cor}$ (most compact IDW) in Eqn. \ref{eqn:coldens1}, adopting $f=0.2$ as before, and setting $N_H$ to $10^{21}$~cm$^{-2}$, yielding:
\begin{equation}
\label{eqn:thresh}
\dot{M}_{\rm thresh} = 1.5 \times 10^{-10} M_{\odot} {\rm yr}^{-1} P^{1/2} R_{\star}^{1/4} M_{\star}^{1/4} \left(T_{\star}/4000\right)^{1/4}  
.\end{equation}
Figure \ref{fig:threshold} compares this $M_{\star}$-dependent threshold for two values of $P_{\star}$ and a range of ages to \mdotacc\ values compiled by \citet{Manara2023}.  Rather than minimum \mdotacc\ values reflecting the rate of PE driven by EUV radiation from the star \citep{Thanathibodee2023}, we interpret this threshold as the minimum \mdotacc\ at which an IDW can quench X-ray-driven PE \citep[see also ][]{Ercolano2023}.      

\subsection{Accretion variability from a limit cycle}
\label{sec:limitcycle}
Feedback between accretion in the inner disk, the attenuation of stellar X-rays, and PE, combined with a delay between the removal of mass by PE and inner disk accretion (i.e., the viscous transport time) raises the possibility of an oscillatory limit cycle \citep{Strogatz2000} that produces variability on that delay timescale.  We demonstrated this phenomenon using a minimalist ``toy" model that captures these feedbacks (rather than the complex model described in Sect. \ref{sec:model}) and that is represented by time-dependent equations for the inner disk accretion and PE rates:
\begin{subequations}
\begin{gather}
\dot{M}_{\rm acc}(t) = \dot{M}_0 - \dot{M}_{\rm PE}\left(t - \Delta t\right), \label{eqn:toy1}\\
\dot{M}_{\rm PE}(t) = \dot{M}_{\rm PE}^0 \exp \left[-a f \dot{M}_{\rm acc}(t) \delta\left(\dot{M}_{\rm acc}(t) < \dot{M}_{\rm crit}\right)\right].\label{eqn:toy2}
\end{gather}
\end{subequations}
Equation \ref{eqn:toy1} describes the decrease in inner disk accretion at time, $t$, from a baseline rate, $\dot{M}_0$, due to PE at time $t - \Delta t$ , with $\Delta t$ representing the transport time (e.g., viscous timescale) from the location of the PE wind to the inner disk edge.   Equation \ref{eqn:toy2} describes the exponential attenuation of X-ray-driven PE below an unattenuated rate, $\dot{M}_{PE}^0$, if an IDW is present (a condition represented by the boolean integer, $\delta$).  The optical depth of the IDW to X-rays is assumed to scale with IDW mass flow times a fixed attenuation coefficient, $a$, the IDW mass flow is \mdotacc\ times a fixed ejection/accretion ratio, $f$, and PE is assumed to scale linearly with X-ray irradiation.   

\begin{figure}[h!]
    \centering
    \includegraphics[width=0.49\textwidth]{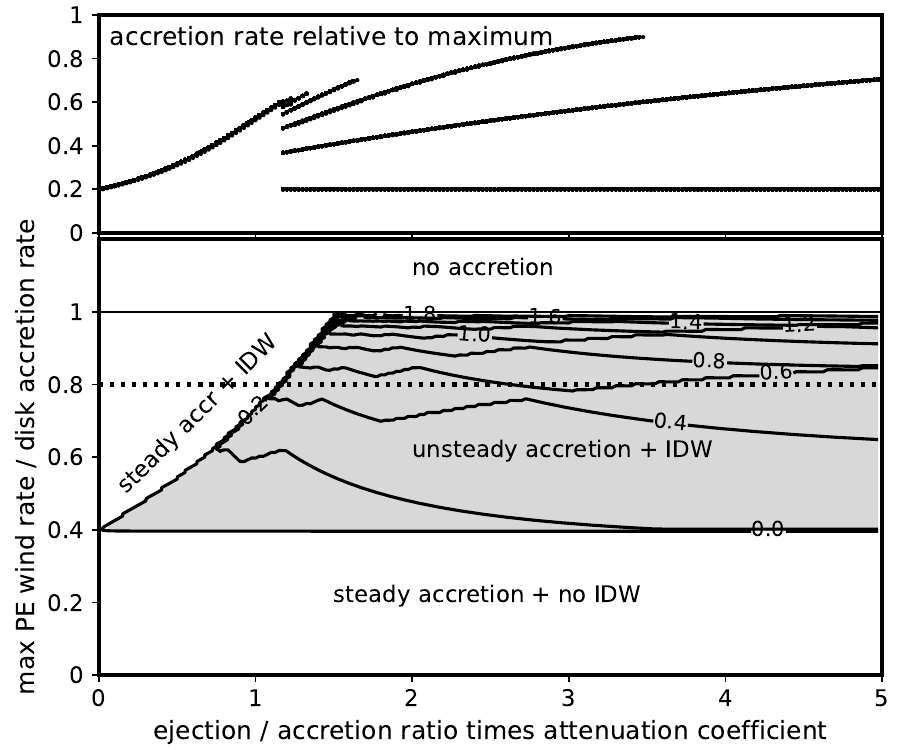}
    \caption{\textbf{Panel (a):} Above a critical value of the product of the X-ray attenuation coefficient, $a$, times the ejection/accretion ratio, $f$ (Eqn. \ref{eqn:toy2}), solutions for the inner disk accretion rate multifurcate.  \textbf{Panel (b):} Domains of solutions to Eqns. \ref{eqn:toy1} and \ref{eqn:toy2} in $a \dot f$ vs. the maximum PE rate.  Contours are the ratio of the maximum to minimum values of \mdotacc\ expressed in dex.  The dotted line is the parameter range for the case shown in the top panel.}
    \label{fig:limitcycle}
\end{figure}

The equations are iteratively solved, with each iteration representing an elapsed time, $\Delta t$.  If the feedback is sufficiently strong -- that is, if $\dot{M}_{PE}^0 > \dot{M}_0 - \dot{M}_{\rm crit}$ and the product $a \times f$ exceeds a threshold value -- the solutions of Eqns. \ref{eqn:toy1} and \ref{eqn:toy2} become multivalued or ``multifurcate" (Fig. \ref{fig:limitcycle}a).  Physically, this could manifest itself as unsteady accretion over the accretion timescale, $\Delta t$, of the inner disk.  Based on the observed dispersion of T Tauri disk masses and accretion rates \citep[e.g,][]{Almendros-Abad2024}, $\Delta t$ could be $\sim10^4-10^5$ yr.   Figure \ref{fig:limitcycle}b shows the domain of multifurcated solutions if $\dot{M}_{\rm crit}$ is 60\% of $\dot{M}_0$.  For \mdotacc\ $\in$ [$0.4\dot{M}_0$,$\dot{M}_0$] and high $a \times f$, there is a regime of unsteady accretion (shaded gray region in Fig. \ref{fig:limitcycle}b with contours representing the ratio of the maximum to minimum values in dex).  For low $a \times f$, \mdotacc\ is steady but attenuated by an IDW.  if $\dot{M}_{PE}^0 < 0.4\dot{M}_0$, there is no IDW and \mdotacc\ = $\dot{M}_0$.  If $\dot{M}_{PE}^0 >\dot{M}_0$, PE completely halts accretion.  The unsteady mode dominates as $\dot{M}_0$ approaches $\dot{M}_{\rm crit}$.  Variation in \mdotacc\ on timescales of $\sim10^4-10^5$ yr, in addition to shorter-term variability \citep[e.g.,][]{Claes2022} could explain some of the observed two-orders-of-magnitude scatter around the relation between \mdotacc\ and $M_{\star}$ \citep{Manara2023}.   

\subsection{Disk lifetime and chemistry}
\label{sec:chemistry}
Infrared spectroscopy has revealed gas chemistry indicative of a super-solar C/O ($\sim 1$) in the inner disks of some VLMSs \citep{Najita2013,Tabone2023,Arabhavi2024,Kanwar2024}, but not all \citep{Xie2023}.  This difference between VLMS and TTS disks has been explained as the result of a more compact snowline, a shorter viscous timescale, and the faster inward transport of gas depleted in oxygen by the condensation of \water- and CO$_2$-rich solids \citep{Mah2023}.  However, it could also be explained by more pervasive suppression of PE that would otherwise form a gap and inhibit later inward transport of O-depleted gas from the outer disk \citep{Gasman2023}.  Of four VLMS disks studied with \jwst\ and compared by \cite{Kanwar2024}, all have low C/O chemistry except the one (Sz\,124) with the highest \mdotacc\ and apparent $L_X$, suggesting a lack of an X-ray attenuating IDW.  Sz\,124 is also the only one with detectable [\ion{Ne}{II}] emission indicative of X-ray-driven PE, perhaps sufficient to open a gap such as the one suggested by the highest-resolution analysis \citep{Jennings2022} and inhibit later transport of O-poor gas to the inner disk. 

\subsection{Disk structure and planet formation} 
\label{sec:planets}
\begin{figure*}[h!]
\centering
\includegraphics[width=0.45\textwidth]{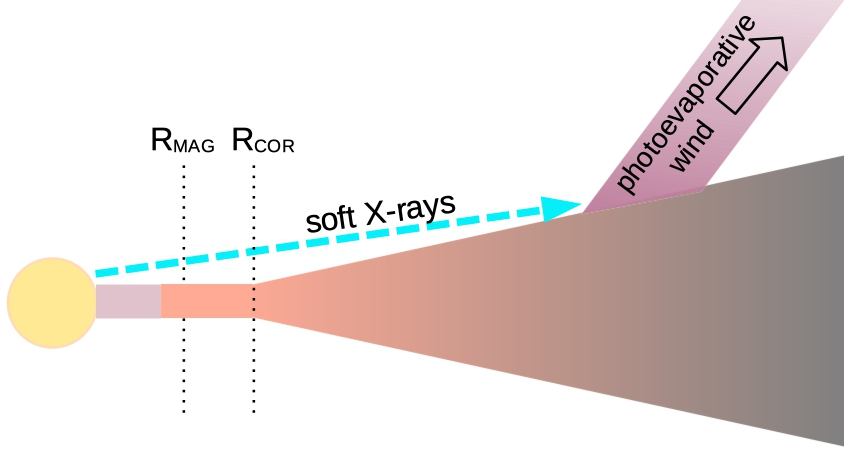}
\hfill
\includegraphics[width=0.45\textwidth]{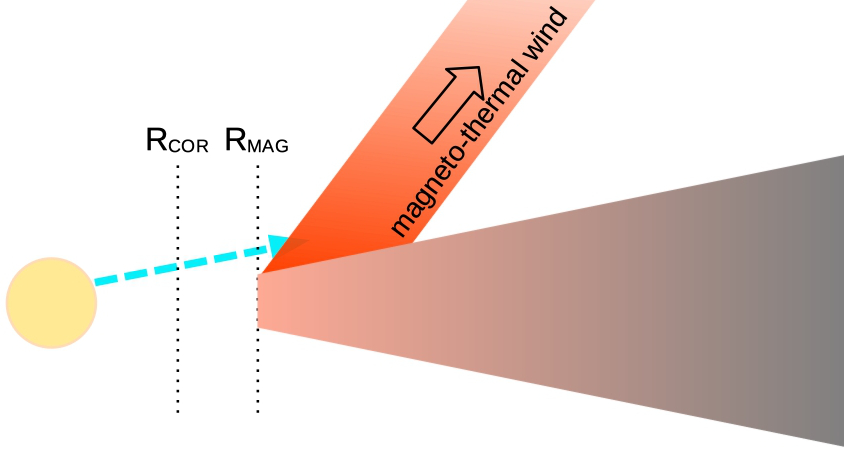}
\includegraphics[width=0.45\textwidth]{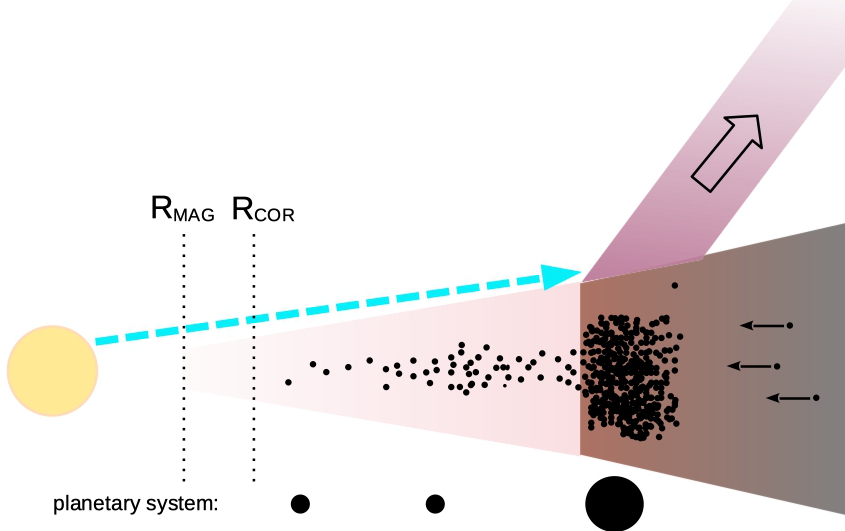}
\hfill
\includegraphics[width=0.45\textwidth]{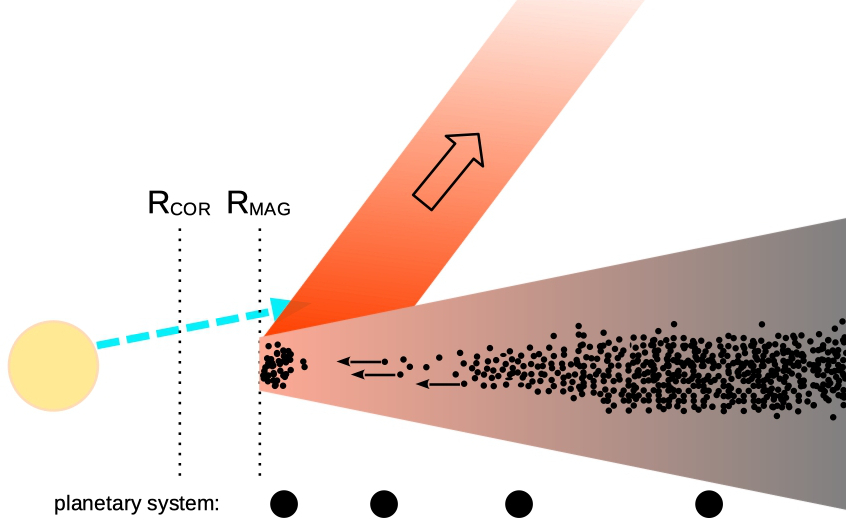}
    \caption{Illustration of two scenarios of disk evolution and planet formation. \textbf{Left:} If $R_{\rm mag} < R_{\rm cor}$, unimpeded X-ray irradiation by the central star opens a gap in the disk that eventually depletes the inner disk and stymies planet formation there.  A pressure bump outside the cavity can trap solids and enable the formation of one or more giant planets.  \textbf{Right:} If $R_{\rm mag} > R_{\rm cor}$, an IDW  shields the disk further out from X-rays, slowing or halting PE and delaying or preventing the opening of a gap.  Solids drift accumulate closer to the star, leading to more compact dust disks and close-in multi-planet systems.}
    \label{fig:cartoons}
\end{figure*}

Diverging disk evolutionary pathways could end in markedly different planetary outcomes (Fig. \ref{fig:cartoons}).  If PE opens a gap early in the disk history,a typical scenario among solar-mass stars (Fig. \ref{fig:evolution3}), a pressure maximum immediately exterior to the gap can trap migrating solids; that is, millimeter-to-centimeter-sized ``pebbles" \citep[e.g.,][]{Pinilla2021}.  This starves the inner disk of condensible solids, especially volatiles \citep{Kalyaan2021}, while accumulation in the pressure maximum could cause the formation of massive cores and giant planets \citep[e.g.,][]{Lau2022}.  A statistical link between transition disks -- a possible stage after gap opening -- and giant planets has been previously described \citep{vanderMarel2021}.  Furthermore, the inward migration of giant planets formed further out could be slowed or halted by the evacuation of the inner disk \citep{Alexander2012b,Monsch2021}.\footnote{Embedded giant planets could also influence the opening of a gap by a PE wind \citep{Weber2024}.}  In longer-lived disks where gap opening is delayed, a likely outcome around lower-mass stars (Fig. \ref{fig:evolution3}), solids will drift unimpeded close to the star \citep{Pinilla2020}, explaining why the continuum emission from dust VLMS disks tends to be more compact \citep{Ansdell2018b,Kurtovic2021}.  Planet formation and migration in denser compact disks could yield multiple short-period planets in near-resonant chains, which are observed around M dwarfs more frequently than solar-type stars \citep{Gaidos2016,Hardegree-Ullman2019,Ment2023}.

\subsection{Caveats}
\label{sec:uncertainties}

Sensitivity to key parameters is investigated in Appendix \ref{sec:sensitivity}. A disk's outcome depends on the stellar period (Appendix \ref{sec:rototation_sensitivity}), the radius (Appendix \ref{sec:stellar_model}), and the time-dependence of \mdotacc\ (Appendix \ref{sec:accretion_sensitivity}), as well as $B_{\star}$.  We emphasize that while this sensitivity makes exact predictions more difficult, it is the underlying mechanism by which diversity in disk evolution can arise.  In the analytic model, we prescribe the values or time-dependence of these parameters, but it is derived self-consistently in our numerical modeling (Appendix \ref{sec:numerical}).  The degree of shielding of the disk from X-rays for a given \mdotacc\ is sensitive to the poorly known geometry and kinematics of the IDW and ejection-accretion ratio (see Appendix \ref{sec:ejection_sensitivity}).  The sonic speed at the base of the wind could vary by a factor of three for temperatures between $10^3$ and $10^4$K.  Shielding could also occur from a more static wall or envelope supported by magnetic pressure \citep{Zhu2024}. The PE rates are based entirely on 2D wind models and have only been tested by observations in a limited way \citep[e.g.,][]{Ercolano2021}.  A model that includes detailed chemistry suggests that PE rates could be lower due to more efficient cooling \citep{Sellek2024b}.  

\section{Conclusions}
\label{sec:summary}
Photoevaporation driven by soft X-rays from the central star is a pathway by which PPDs develop gaps and inner cavities and eventually dissipate.  An IDW or other vertical structure near the inner edge of a disk can shield the outer disk from X-rays, suppressing PE and extending its lifetime.  Such a wind is present when the partially ionized disk is truncated by the stellar magnetic field beyond the corotation radius, but will disappear when the contraction of the protostar, its field, and the truncation radius outpaces the effect of a declining accretion rate, allowing X-rays to reach the disk.  The timing of these events depends sensitively on the stellar mass, large-scale dipole field strength, and rotation.  An analytic model of star and disk evolution shows that this inner-outer disk connection explains:
\begin{itemize}
    \item the observed trend of increased disk lifetime and decreased occurrence of disk structure around lower-mass stars;
    \item the 1 dex range in disk lifetime for a given mass, with rapidly rotating stars with stronger magnetic fields having longer disk lifetimes; 
    \item the paucity of disks with accretion rates lower than $\sim10^{-10}$ \msun\,yr$^{-1}$, as below that level the IDW is too diffuse to significantly attenuate X-rays and prevent disk dispersal;
    \item some of the observed scatter in disk accretion rates may be a manifestation of kiloyear-timescale variability, specifically a limit cycle due to feedback between PE, inner disk accretion, and the attenuation of X-rays by an accretion-fed wind.
\end{itemize}
A dichotomous outcome in disk evolution suggests a corresponding divergence in disk chemistry and planet formation outcomes.  Disks with early-developing gaps and cavities tend to generate giant planets on 1--10 au orbits.   Longer-lived disks where suppression of PE has delayed gap formation are more likely to have higher C/O due to the unimpeded inward transport of O-poor gas, and spawn systems of multiple close-in super-Earths and Neptunes.

Direct evidence that the attenuation of X-rays and suppression of PE prolongs disk life is needed. \citet{laos2022} found that 0.8--3 keV X-ray emission from six M dwarf stars hosting long-lived disks is not systematically lower than that from their diskless coeval counterparts.  However, this energy range is less sensitive to attenuation by gas and less relevant to PE.  Moreover, if the wind has a wide opening angle \citep[e.g., $\sim$60\,deg,][]{Nemer2024} then attenuation would not be observed along our line of sight among half of the disks in a randomly oriented sample.  Observations of more long-lived disks at softer energies ($<1$\,keV) are warranted.  Key to more robust tests of this paradigm are near-simultaneous measurements of the large-scale magnetic field strength, accretion rate, and winds as might be accomplished with synergistic combinations of ground- and space-based spectroscopy in the UV and infrared.

\begin{acknowledgements}
The authors thank Kees Dullemond for a thoughtful and careful review, and Ilaria Pascucci for helpful comments.  EG was supported as a Gauss Professor at the University of G\"{o}ttingen by the Nieders\"{a}chsische Akaemie der Wissenschaften.
\end{acknowledgements}

\begin{appendix}

\section{Numerical modeling}
\label{sec:numerical}

We compared the analytic results represented in Fig.~\ref{fig:evolution2} to the output of a numerical model that combines stellar and disk evolution self-consistently.  The implicit hydrodynamic TAPIR code \citep[e.g.,][]{Ragossnig2020} includes a detailed calculation of the disk truncation radius \citep[][]{Steiner2021} and a stellar spin model \citep{Gehrig2022}, and calculates photoevaporative mass loss from the disk \citep[][]{Cecil2024}.  

There are several key differences between the numerical and analytic models.  First, the stellar rotation period is not fixed but evolves according to changes in moment of inertia and torques from the disk \citep{Gehrig2022}.  Second, the code calculates the truncation radius based on the stellar magnetic pressure, the ram pressure of the infalling material, and the static gas pressure in the innermost disk region \citep[e.g.,][]{Romanova2002, Bessolaz2008, Steiner2021}.  Finally, disk accretion is calculated continuously and self-consistently with an $\alpha$ viscosity, a ``dead" zone, and PE as a sink term.  Feedbacks between PE, inner disk accretion, the location of the magnetic truncation radius, the operation of an IDW, and the attenuation of stellar X-rays are included.  We use the model as described in \citet{Cecil2024}, but included a ``switch" that turns off PE when $\mathcal{R}$ reaches unity.  To avoid numerical difficulties, PE is reduced smoothly starting at $\mathcal{R}=0.9$ to zero at $\mathcal{R}=0.9$.  

In all runs, the $\alpha$ viscosity parameter is taken to be $5 \times 10^{-3}$ ($2 \times 10^{-4}$) outside of (within) the dead zone.  In these simulations, disk lifetime refers to the time at which \mdotacc{} falls below $10^{-11}$ \msun\,yr$^{-1}$.  We assume an initial stellar age of 1~Myr, and stellar radii and luminosities are taken from \citet{Baraffe2015}.  The initial accretion rates are $1.3\times10^{-9}~\mathrm{M_\odot/yr}$ and $7.7\times10^{-9}~\mathrm{M_\odot/yr}$ and the X-ray luminosities (in ergs\,sec$^{-1}$) are taken to be $\log L_\mathrm{X} = 29.5$ and 30.3  for 0.3 and 1.0~\msun\ stars, respectively.  

\begin{figure}
    \centering
    \includegraphics[width=0.5\textwidth]{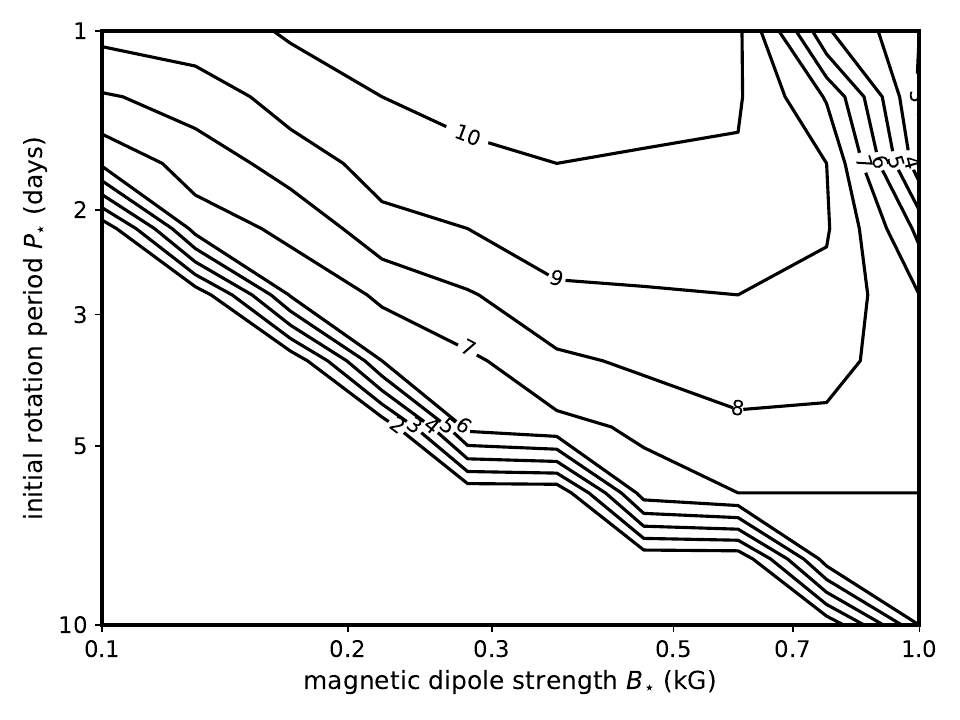}
    \caption{Disk lifetime vs. $B_{\star}$ and initial (not fixed) $P_{\star}$ for a 1-\msun\ star predicted by the numerical TAPIR star-disk interaction model.  Compare to Fig. \ref{fig:evolution2}.}
    \label{fig:tapir_1.0}
\end{figure}

\begin{figure}
    \centering
    \includegraphics[width=0.5\textwidth]{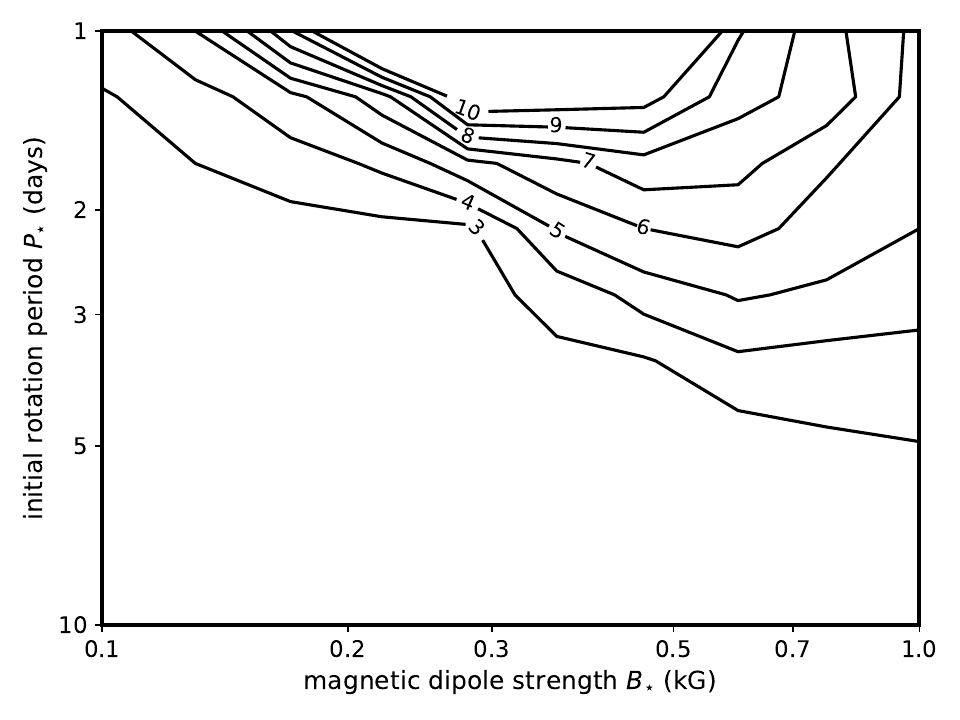}
    \caption{Same as Fig. \ref{fig:tapir_1.0} but for a 0.3-\msun\ star.}
    \label{fig:tapir_0.3}
\end{figure}

Figures~\ref{fig:tapir_1.0} and \ref{fig:tapir_0.3} plot disk lifetime predicted by the model vs. $B_{\star}$ and initial (not fixed) $P_{\star}$ for 1.0 and 0.3\msun\ stars, respectively.  These TAPIR results support the overall prediction of the analytic model that PE is suppressed and disk lifetimes are longer around more rapidly rotating stars with strong magnetic fields (compare to Fig. \ref{fig:evolution2}), and that this effect is more pervasive around stars with lower masses (compare to Fig. \ref{fig:evolution3}).  Differences between the numerical and analytic predictions are attributable to the addition of stellar spin evolution in the former.  For a given magnetic field strength, stellar rotation tends to evolve toward an equilibrium state that is different from the initial condition \citep[e.g.,][]{Ireland2021, Gehrig2022,Serna2024}.  One notable difference is \emph{shorter} disk lifetime for stars with the strongest $B_{\star}$ and shortest initial $P_{\star}$ (upper right-hand corner of Figs. \ref{fig:tapir_1.0}--\ref{fig:tapir_0.3}).  This is a result of a stronger field exerting torques on both the star and disk, enhancing accretion towards the star and moving the disk truncation radius inward, while spinning down the star and moving the corotation radius outward.  This tends to suppress the formation of an IDW which would otherwise protect the disk from PE.  However, the condition of both rapid rotation and a strong field would probably be transient in the Class I phase and not a realistic state at the beginning of the Class II phase.

\section{Sensitivity analyses}
\label{sec:sensitivity}

\subsection{Rotational evolution} 
\label{sec:rototation_sensitivity}

\begin{figure}
    \centering
    \includegraphics[width=0.5\textwidth]{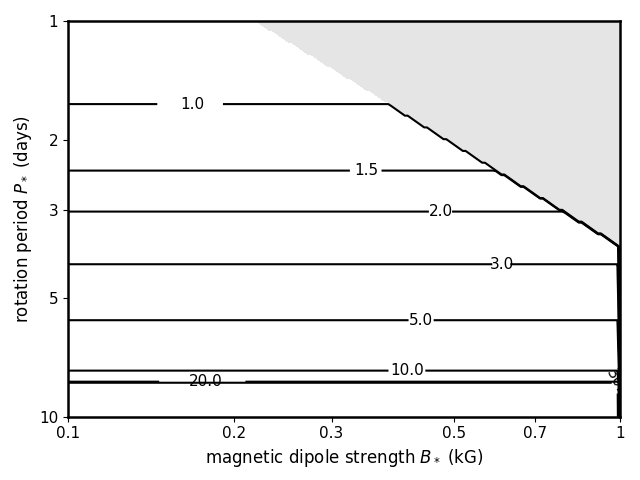}
    \includegraphics[width=0.5\textwidth]{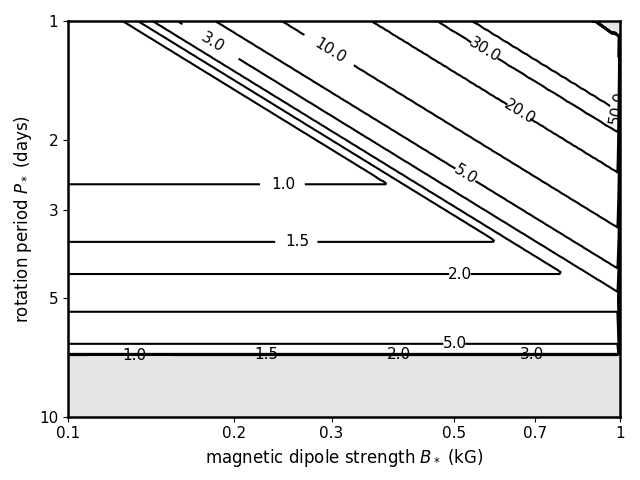}
    \caption{$t_{\rm gap}$ vs. $B_{\star}$ and $P_{\star}$ for 1\,\msun.  Same as Fig. \ref{fig:evolution2} except using the stellar magnetosphere \citep{Koenigl1991} (top) and trapped flux \citep{Ostriker1995} models to calculate evolution of $P_{\star}$.  The ordinate is now the minimum rotation period over the disk lifetime.}
    \label{fig:rotation}
\end{figure}

Models of star-disk interaction predict that rotation periods do not remain constant as assumed in our analytical model (Sect. \ref{sec:model}), but instead that they evolve slightly over 1--10 Myr.  The ``stellar magnetospheric" (large-scale dipole; \citealt{Koenigl1991}) and ``trapped magnetic flux" models of star-disk interaction \citep{Ostriker1995,Johns-Krull2002} predict rotation periods that scale as $P_{\star} \propto R_{\star}^{18/7}\dot{M}^{-3/7}$ and $P_{\star} \propto R_{\star}^{4}\dot{M}^{-1}$, respectively.  (See \citealt{Mueller2024} for an observational test of these models.). We re-calculated gap opening times for the solar-mass case using these relations.  Figure \ref{fig:rotation} plots these results, where the ordinate is now the minimum value of $P_{\star}$ that occurs during the simulation.  The overall distribution of gap opening times is similar (compare to Fig. \ref{fig:evolution2}) except that in the trapped flux scenario, late (rather than no) gap opening is predicted for rapidly rotating stars with high $B_{\star}$, and no gap opens for very slow rotators.  We also find that the scaling relations do not predict an interval of quasi-constant rotation for the 0.3\,\msun\ case, which could reflect either reality or the limitations of our prescription for the accretion rate history of VLMSs.

\subsection{Stellar model}
\label{sec:stellar_model}

\begin{figure}
    \centering
    \includegraphics[width=0.5\textwidth]{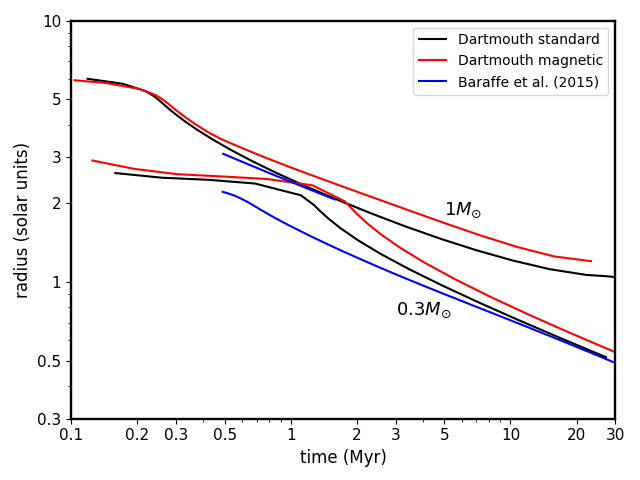}
    \caption{Stellar radius evolution for 1- and 0.3-\msun stars predicted by the Dartmouth standard model \citep{Dotter2016}, the Dartmouth magnetic model , which includes the inflationary effect of magnetic fields via spots and inhibition of convection \citep{Feiden2016}, and the \citet{Baraffe2015} model.  In the Dartmouth models a different set of abundances and opacities results in prolonged deuterium burning compared to the \citet{Baraffe2015} model (G. Feiden, pers. comm.).  This delays stellar contraction and potential affects disk evolution via the radius dependence of Eqn. \ref{eqn:mdotcrit}.}
    \label{fig:stellarmodel}
\end{figure}

For a given stellar mass, rotation period, and dipole magnetic field strength, the critical value of accretion rate at which an IDW appears is highly sensitive to stellar radius (Eqn. \ref{eqn:mdotcrit}).  Predictions of stellar radius by models depend on the physics that is included (Fig. \ref{fig:stellarmodel}).  Deuterium burning increases the entropy and can delay the early contraction of pre-MS stars \citep{Stahler1998}.  The timescale of D-burning in fully convective pre-MS stars is sensitive to the central temperature, which in turn depends on the adopted abundances and opacities (compare Darmouth and \citet{Baraffe2015} models in Fig. \ref{fig:stellarmodel}).  In most cases, deuterium burning occurs too early to matter, with the possible exception of slowly rotating M dwarfs, where it is not completely burned for 3 Myr (Fig. \ref{fig:stellarmodel}).  For these cases it could prolong the disk by delaying contraction and increasing $\dot{M}_{\rm crit}$.  

Internal magnetic fields can significantly inflate stellar radii over a range of masses and ages via suppression of convection or spots \citep{Feiden2016}.  This would increase $\dot{M}_{\rm crit}$ and, all else being equal, prolong the lifetime of disks, except for slow rotators where stellar X-ray emission is the limiting factor (e.g., Fig. \ref{fig:evolution}c).  

\subsection{Accretion rate}
\label{sec:accretion_sensitivity}

\begin{figure}
    \centering
    \includegraphics[width=0.5\textwidth]{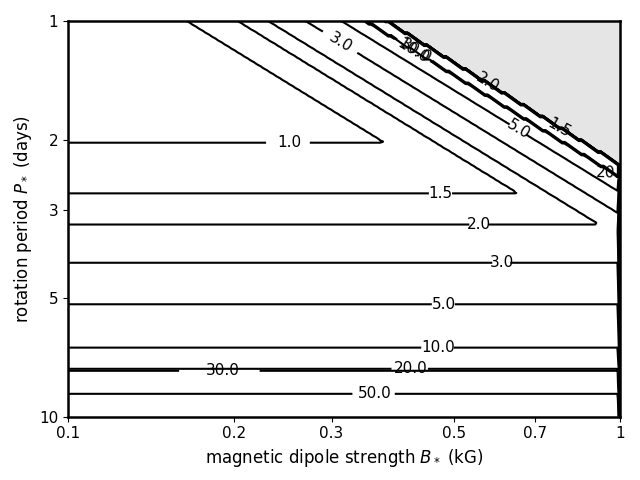}
     \includegraphics[width=0.5\textwidth]{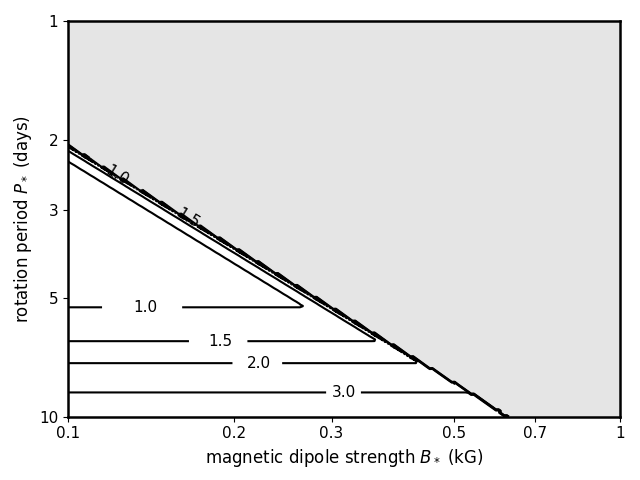}
    \caption{$t_{\rm gap}$ vs. $B_{\star}$ and $P_{\star}$ for 1\,\msun. Same as Fig. \ref{fig:evolution2} except with a trebled initial accretion rate (top) or an accretion rate power law index of -2 instead of -5/4.}
    \label{fig:sens_accretion}
\end{figure}

Whether PE opens a cavity or is attenuated by an IDW depends on the disk accretion history. We demonstrated this sensitively by two departures from the nominal case for a 1\,\msun\ star (Eqn. \ref{eqn:accrete}), first by tripling the initial value \mdotacc$(0)$, and second by changing the power law index from -5/4 to -2.  The index value of -2 corresponds to the case where the disk viscosity depends on radius as $r^{3/2}$, the maximum dependence consistent with observed surface density profiles and disk demographics \citep[an index greater than -5/4 and negative radius dependence is also excluded by these observations][]{Alexander2023}.  These changes slightly decrease and significantly increase the $B_{\star}$--$P_{\star}$ parameter range where PE is completely suppressed, respectively (top and bottom panels of Fig. \ref{fig:sens_accretion})   

\subsection{Ejection/accretion ratio}
\label{sec:ejection_sensitivity}

\begin{figure}
    \centering
    \includegraphics[width=0.5\textwidth]{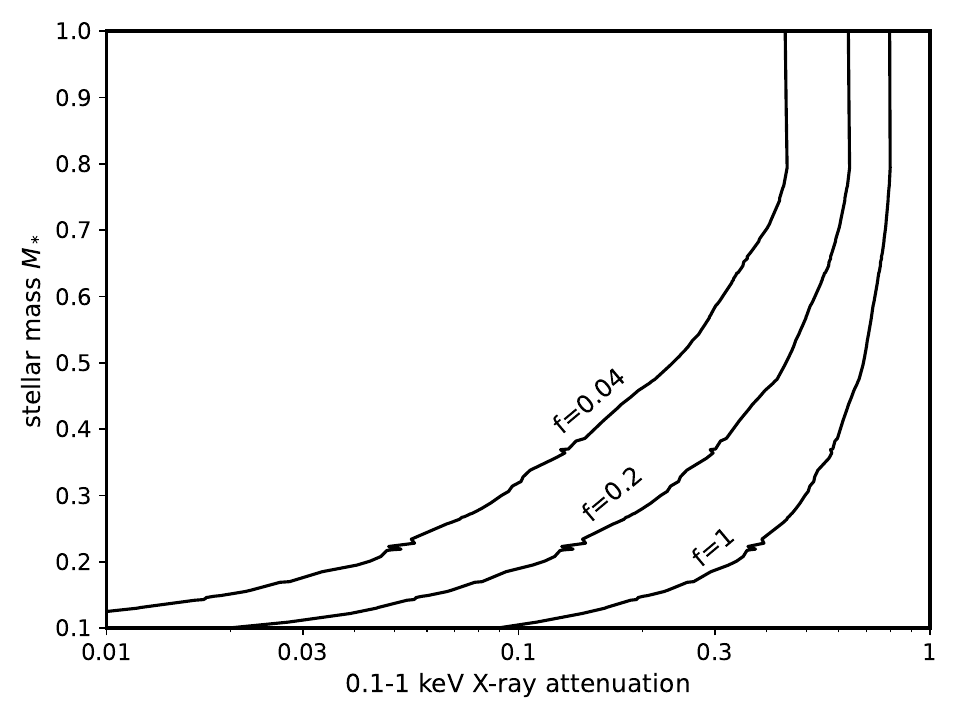}
    \caption{0.1--1 keV X-ray attenuation at $\mathcal{R} = 1$ vs. stellar mass at the end of the IDW phase for different values of ejection/accretion ratio $f$.}
    \label{fig:attenuation}
\end{figure}

For a given accretion rate, the attenuation of X-rays by an IDW depends on the ejection/accretion ratio $f$, which is poorly constrained but thought to range between $\sim$0.01 and 1 \citep{WatsonD2016,Serna2024}.  Figure \ref{fig:attenuation} plots the X-ray attenuation at the end of the IDW phase ($\mathcal{R} = 1$) vs. $M_{\star}$ using Eqn. \ref{eqn:coldens1}, our nominal two-temperature model of coronal X-ray emission, and different values of $f$.   Attenuation is also sensitive to the flow velocity at the foot of the wind (here assumed to be the sound speed) and the angle of the wind with respect to the disk normal (see Sect. \ref{sec:uncertainties}).
\noindent 

\subsection{Stellar coronal X-ray spectrum}

X-ray absorption is strongly energy-dependent, soft photons being absorbed more efficiently than hard photons, and thus the effective attenuation over 0.1--1\,keV by intervening gas depends on the spectrum of the intrinsic emission from the star. We compared attenuation for different coronal spectra typically used for TTSs, i.e., isothermal plasmas with $kT $= 0.5, 0.7, 1.0, and 2.0\,keV \citep[][see our Sect. \ref{sec:model} for details]{Telleschi2007}.  Figure \ref{fig:attenuation} plots the calculated attenuation for single-temperature (0.7 and 2 keV) model plasmas, plus a mix of two plasma components at 0.5~keV plus 1~keV with an emission measure ratio of 2:1.  The effective attenuation depends only weakly on the temperature because the different spectra have a similar shape over 0.1--1\,keV. This is a consequence of the spectra being dominated by a flat bremsstrahlung continuum plus widely distributed emission lines.  The denser line forest between 0.7--1~keV is present and similar for all spectra in this temperature range.  The strongly temperature-dependent turnover of the bremsstrahlung spectrum from flat to exponentially decreasing occurs around $kT$ and therefore affects spectra at energies above 1~keV, a range not relevant to PE.

\begin{figure}  
    \centering
    \includegraphics[width=0.5\textwidth]{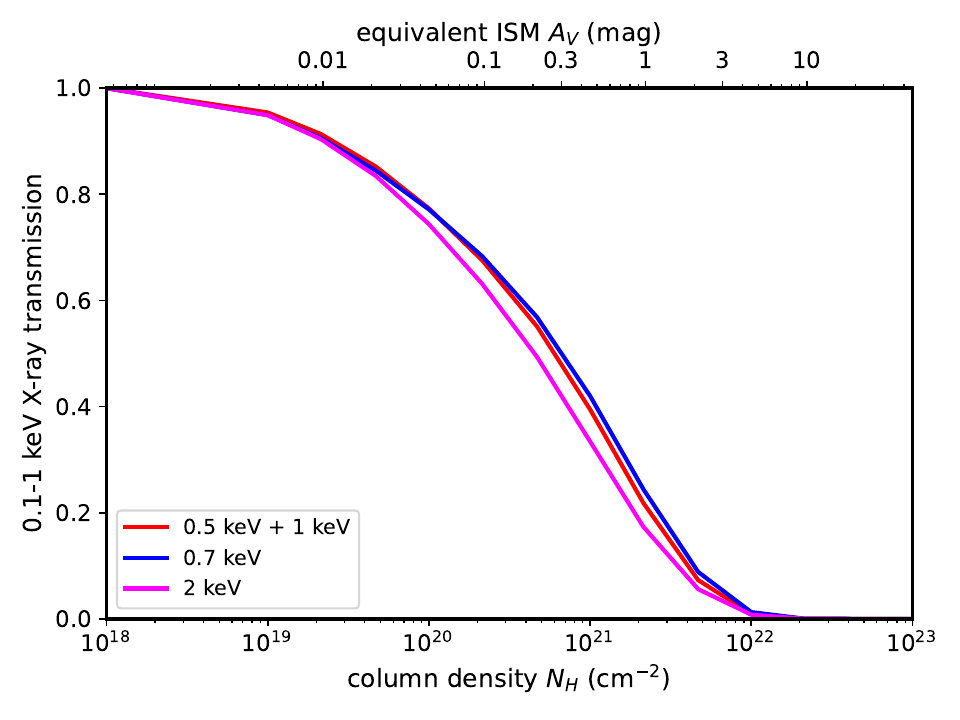}
    \caption{Attenuation of 0.1--1\,keV X-rays for different coronal emission models.}
    \label{fig:attnuation}
\end{figure}

\end{appendix}

\end{document}